\documentclass[conference,10pt]{IEEEtran}
\IEEEoverridecommandlockouts

\usepackage{color}

\ifCLASSOPTIONcompsoc
  \usepackage[nocompress]{cite}
\else
  \usepackage{cite}
\fi

\ifCLASSINFOpdf
\else
\fi
\usepackage{amssymb}
\usepackage[cmex10]{amsmath}
\usepackage{amsthm}
\usepackage{bm}
\usepackage{algorithm}
\usepackage{algorithmic}
\usepackage{graphicx}
\usepackage{epstopdf}
\usepackage{subfigure}
\usepackage{booktabs}
\usepackage{cite}
\usepackage{color}
\usepackage{comment}

\newtheorem{lemma}{Lemma}
\newtheorem{theorem}{Theorem}

\newtheorem{assumption}{Assumption}
\newtheorem{remark}{Remark}

\begin{document}

\title{EntrapNet: a Blockchain-Based Verification Protocol for Trustless Computing\\
}

\author{Chong~Li, \textit{Senior Member, IEEE,} 
        Lei~Zhang,
        and~Senbiao~Fang% <-this % stops a space
\IEEEcompsocitemizethanks{\IEEEcompsocthanksitem The authors are with the Department of Electical Engineering, Columbia University, 
New York, USA, 10036.\protect\\
E-mail: \{cl3607, lz2671,sf2977\}@columbia.edu
}}

\maketitle

\begin{abstract}
In this paper, we propose a blockchain-based computing verification protocol, called \textit{EntrapNet}, for distributed shared computing networks, an emerging underlying network for many internet of things (IoT) applications. EntrapNet borrows the idea from the practice of entrapment in criminal law to reduce the possibility of receiving incorrect computing results from trustless service providers who have offered the computing resources. Furthermore, we mathematically optimize EntrapNet to deal with the fundamental tradeoff of a network: security and efficiency. We present an asymptotic optimal solution to this optimization. It will be seen that EntrapNet can be performed as an independent and low-cost layer atop of any trustless network that requires outsourced computing, thus making secure computing affordable and practical.
\end{abstract}


\section{Introduction}
Like electricity, nowadays computing power is an essential utility in human's daily life, ranging from education, science to marketing and media. Recently, many real-world problems are being reshaped by mathematically analyzing big data. This trend leads to a stronger need for computing power. However, access to powerful computing devices (e.g., cloud servers) has always been unfair and undemocratic as the cost of using these computing devices is typically high. On the other hand, the waste of computing power from billions of idle devices in families and data centers around the globe is tremendous. Thus, how to utilize these idle computing devices to satisfy the growing need of computing power becomes an intriguing and challenging research topic.  Along the line, blockchain technology is promising in the sense that these idle geo-distributed computing devices can be tied together by a blockchain network, a server-less and secure network (e.g., projects \cite{canonchain, golem, sonm, zcash}). Specifically, a client/user on the network can rent the computing power from a computing power provider (\textit{provider} in short) anywhere in the world to help complete one or more computing tasks. The task can be machine learning training, 3D rendering, scientific computation and more. The client then pays the service with cryptocurrency like bitcoin and the provider who has offered computing resources is rewarded in return. In essence, this can be treated as a distributed shared computing platform or marketplace with the objective to fully utilize the computing resources around the globe.  

However, a long-standing problem of outsourcing computational tasks to another party is how a client/user can verify the result efficiently without re-executing the task. On one hand, providers do not necessarily have strong incentives to ensure correctness. On the other hand,  for complex and large-scale providers (e.g., cloud servers). it is unlikely to guarantee that the execution is always correct due to mis-configurations, randomness in hardware and more. This problem, recognized as \textit{verifiable computing}, has been studied by computer scientists for nearly a decade. 

\subsection{Related Work}
A straightforward solution to this problem is to replicate computations on multiple computing devices \cite{multiServer01,multiServer02,multiServer03}. However, this solution implicitly assumes that failures from these computing devices are uncorrelated. This assumption can be invalid in many cases, for example, cloud servers always have homogeneous hardware and software platforms. Another solution is based on the method of taking a small group of samples and then auditing the responses in these samples. However, if the incorrect outputs do not occur frequently, this solution may not perform well. One can also find other solutions such as attestation \cite{Bootstrapping} and trusted hardware \cite{trustHardware}, but these solutions always require a chain of trust and guaranteed correct hardware computation. 
One groundbreaking approach that uses the traditional technique is to force the provider to provide a \textit{short proof} which can prove the correctness of the computation. This short proof should be ``short and easy-to-check'' compared to re-executing the computing task.  Many proof-based systems have been developed recently, including Pinocchio \cite{Pinocchio}, TinyRAM \cite{TinyRAM}, Pantry \cite{Pantry} and Buffet \cite{Buffet}. However, due to the tremendous overhead of the cryptographic setup between the client and provider, and the computational cost for the provider to construct such a short proof, these systems are only near practical \cite{survey_walfish}.
Another novel approach leverages the blockchain technology to secure outsourced computation. One example is TrueBit \cite{TrueBit}, a smart contract that verifies the computational results through a trustless economic protocol. However, the 5-to-50 times more cost of using Truebit and its restriction to Ethereum network could be the bottleneck for its massive adoption. 
\textcolor[rgb]{0,0,0}{In addition, a new approach that has similar idea with our work is to add some precomputed subtasks in original data set and verify them upon task completion. For example, \cite{biometric} has presented a mechanism to verify the outsourced computation for biometric data, which inserts some precomputed fake items into biometric data set to detect misbehaviour of lazy servers. However, this method only focused on specific computation algorithms and data set structures. Furthermore, inserting fake items into every computation task for every server is lack of efficiency.}

In this paper, we introduce a novel verifiable computing protocol, called \textit{EntrapNet}, which borrows the idea from the practice of criminal entrapment. EntrapNet can be viewed as an independent blockchain-based computing verification layer atop of a distributed shared-computing network. Therefore, it can be easily implemented to protect any IoT network that invokes outsourced computing. The paper is organized as follows. In Section \ref{sec::arch}, we present the overall system architecture and describe the details of each key building block in EntrapNet. In Section \ref{sec::opt}, we propose a general mathematical framework to optimize EntrapNet and then provide asymptotically optimal solutions.

{\bf Notations:} Uppercase and the corresponding lowercase letters, e.g., $(Y,Z,y,z)$, denote random variables and their realizations, respectively. $Pois(x;\lambda_x)$ represents the distribution of a Poisson random variable $X$ with parameter (or mean value) $\lambda_x$. We use log to denote the logarithm with base 10.

\section{System Architecture}\label{sec::arch}
In this section, we review the EntrapNet protocol. We start out with an overview on the system architecture and then dive into the details on each key aspect of the system. First of all, EntrapNet makes two basic assumptions as follows:
\begin{assumption}
The blockchain network used in the EntrapNet is reliable.
\end{assumption} 

Note that how a blockchain network remains reliable against cyberattacks is an independent research topic which is out of the scope of this paper and will be studied separately.  

\begin{assumption}
All participants in the network are rational in the sense that their actions are intended to maximize individual profits. 
\end{assumption}

\begin{figure}
\begin{center}
\includegraphics[scale=0.45]{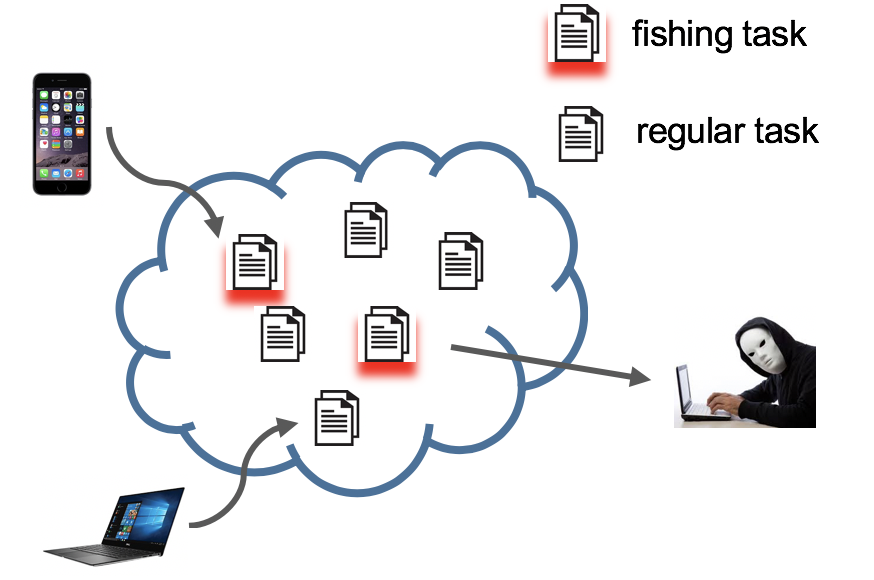}
\caption{EntrapNet: the task pool contains both fishing tasks and regular tasks which the malicious node cannot \textcolor[rgb]{0,0,0}{ distinguish.}}
\label{fig:EntrapNet}
\end{center}
\end{figure}

\subsection{Overview}
In criminal law, entrapment is a practice whereby a law enforcement officer induces a person to commit a criminal offense that the person would have otherwise been unlikely or unwilling to commit \cite{wiki::entrap}. EntrapNet borrows such an idea in which a client/user in the network, being a role of ``officer'', aims to catch a malicious provider in the network by assigning the provider with a fishing task. Since the outcome of the fishing task is predictable/known in advance by the officer,  any misconduct on computing can be easily detected by the officer. In what follows, we provide an overview on the EntrapNet mechanism, which will be discussed in two phases. Fig. \ref{fig:EntrapNet} shows an overall illustration of EntrapNet.

\textit{Phase 1 - Preparation}: 
For a user who wishes to be an officer, \textcolor[rgb]{0,0,0}{he/she} needs to build up a fishing-task repository.  Such a repository contains one or more computing tasks whose results are known and verified by the network. In addition, the information of each fishing task in the repository is written to the \textcolor[rgb]{0,0,0}{repository smart contract on blockchain, which will be activated later for the process of result verification.} Note that based on the incentive protocol \textcolor[rgb]{0,0,0}{(see Section \ref{incentive protocol})}, each user can choose to be an officer or not. Meanwhile, a relatively large amount of deposit is required for a provider to participant in offering its idle computing resources, \textcolor[rgb]{0,0,0}{which will be sent to the incentive pool on blockchain. The incentive pool is another smart contract in charge of storing deposit from providers and paying reward to officers}. It is noteworthy that such a deposit is always far more than the reward received by the provider for correctly completing a computing task. By doing so, the provider would lose big if caught to be faulty. 

\begin{figure}
\begin{center}
\includegraphics[scale=0.5]{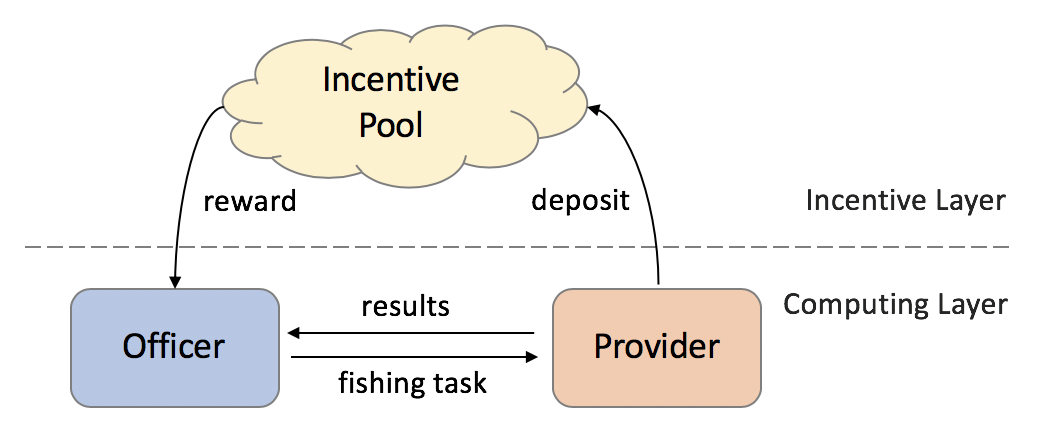}
\caption{EntrapNet Incentive Protocol.}
\label{fig:sys}
\end{center}
\end{figure}

\textit{Phase 2 - Execution}:
In this phase, the network randomly schedules a task from the task pool to a provider, in which the task pool contains both fishing tasks and regular tasks submitted to the network in a fixed time slot, e.g., one minute. \textcolor[rgb]{0,0,0}{The information of all the tasks in the task pool are written into the task pool smart contract on blockchain, the task scheduler will invoke the smart contract that randomly assigns a task to providers.}  Since EntrapNet ensures that a malicious provider cannot filter out the fishing tasks, it is likely for such a provider to execute a fishing task as usual in its faulty manner. Once the provider feeds back the outcome to the officer \textcolor[rgb]{0,0,0}{and record its hash value on blockchain}, the officer can easily identify the correctness of the results by comparing the outcome with the one in its fishing-task repository. If identified to be faulty, the officer post both results on the blockchain via verification smart contract. The judges, a subset or all of the nodes who run this verification smart contract, will rule on whether the officer is correct \textcolor[rgb]{0,0,0}{(e.g. through voting)}. If proved to be correct, the officer will be incentivized with a reward from the EntrapNet incentive pool whereas the deposit of the provider will be forfeited and contributed to the EntrapNet incentive pool. An illustration of the aforementioned mechanism is shown in Fig. \ref{fig:sys}, in which EntrapNet is considered as a combination of two functional layers: incentive layer and computing layer, \textcolor[rgb]{0,0,0}{both consist of several smart contracts on a blockchain system.}

One can see that EntrapNet could be very effective to secure outsourced computing if enforcement (of entrapment) is highly strengthened or the financial punishment on a malicious provider, if caught, is extremely severe. To strengthen the enforcement, the network can simply provide higher reward to officers who successfully catch misconduct. As a consequence, more users could be willing to be officers and submit fishing tasks more frequently. However, the overhead of the network in which many providers compute tasks whose results are already known and have been recorded on blockchain could be much higher, resulting in a more secure but much less efficient system. On the other hand, EntrapNet can sizably increase the required deposit from the provider.  However, a prohibitively high deposit strongly discourage the participation of trustable providers to offer their computing resources. Therefore, it is nontrivial to optimize these hyper-parameters in EntrapNet system \textcolor[rgb]{0,0,0}{and balance the security and efficiency of the system.} In Section \ref{sec::opt}, we will present a general mathematical framework to optimize the system and provide an asymptotically optimal solution.

In the following sections, we will describe the technical details in each EntrapNet building block,  including fishing-task repository, task scheduler, result verification, and incentive protocol, \textcolor[rgb]{0,0,0}{All of them are realized by smart contracts and one smart contract can invoke another.}

\subsection{Fishing-task Repository}\label{sec::fishing_task}
To be an officer, a user first needs to prepare a set of fishing tasks whose computing results need to be verified by the network.  One approach of generating a fishing task (e.g., a Python script) is described as follows. A user first submits a regular computing task {\footnote{In this paper, when submitting a task to the network, we mean that a user is submitting a task request to the blockchain network for a scheduler to schedule this task. Once scheduled, the task script (e.g., Python codes) will be transferred to the provider in a peer-to-peer manner.}} to the computing network along with a flag indicating that this task will be used later as a fishing task. If such a flag is detected by the task scheduler (see Section \ref{sec::task_schedular}), this task will be executed by a set of witnesses in the network.\footnote{A witness is defined as a highly reputable user with a real-world identity, such as banks, conglomerates and \textcolor[rgb]{0,0,0}{big cloud server like Google Cloud.}} These witnesses then send results to the verification smart contract (see Section \ref{sec::res_verfication} for the result verification) \textcolor[rgb]{0,0,0}{and smart contract will check the identity of witnesses. If the identity is correct and these results are proved (by the smart contract) to be matched}, one can claim that the result is verified by the network. At this moment, information of the fishing task will be put into the repository smart contract, and its result will be record in the verification smart contract. the user will receive from the network a proof associated with the task script and the corresponding results. For instance, this proof can be an output of the verification smart contract when the computing results are verified. \textcolor[rgb]{0,0,0}{Also, each fishing task will not be assigned to the same identity twice, in case that this identity in the network becomes familiar with this fishing task.}

\textcolor[rgb]{0,0,0}{Now, the user can add this task to his repository and posts a record into the repository smart contract on blockchain}, indicating that this task in his repository has been verified as a fishing task. In EntrapNet, such a record is an abstract (e.g. a hash via SHA-1) generated from a block containing the following attributes:
\begin{enumerate}
    \item task script;
    \item computing results;
    \item network verification proof;
    \item a user-specific key (e.g., a private key or a field like Bitcoin nonce).
\end{enumerate}
Note that a fishing task in the offer's repository needs to be removed if this task has been used to successfully catch a malicious provider.  This is because, otherwise, this provider can announce this task script on the blockchain, making this fishing task known to every user in the network.  

\subsection{Task Scheduler} \label{sec::task_schedular}

Nowadays there exist many well-studied compromise-based scheduling algorithms, such as proportional fair scheduler widely used in the communication systems. In EntrapNet, however, we suppose that malicious providers are  capable of processing any information to avoid being trapped. In other words, any compromise-based design in scheduling will be utilized by malicious providers to lower the probability of being caught. Therefore, EntrapNet employs a uniform randomly task scheduler. That is, a submitted task is assigned to all providers with equal probability. Another benefit of doing so is that the computation needed in scheduling is minimized. This indicates that such a task scheduler can be easily implemented in a distributed fashion, say, executed by smart contracts in Ethereum.

\begin{remark}
In practice, the smart contract of scheduling can be executed by one node (rather than all nodes in the network) that is dynamically selected in every scheduling interval. For instance, this node can be a witness node chosen according to the rule of round robin among all witness nodes.  Another example is that the node is chosen according to the rule of proof of stake (PoS) \cite{PoS} among the witness nodes in which a witness who possesses $p$ fraction of the total amount of coins among all witnesses becomes eligible to schedule the next task with probability $p$. 
\end{remark}

\textcolor[rgb]{0,0,0}{The Task Scheduler in EntrapNet can obtain a small amount of transaction fee from each scheduled task.} In practice, as the number of task submissions from individuals follows Poisson process over time, it is likely to have no task submitted in a given time slot $T$, i.e., an empty queue for scheduling. To facilitate system analysis and optimization (will be seen in Section \ref{sec::opt}), we make the following assumption.
\begin{assumption}\label{ass:onetask}
In every time slot $T$ there exists at least one task in the queue for scheduling.
\end{assumption}
In other words, EntrapNet always imposes one computing task in the queue in every time slot $T$. To achieve this goal, for example, the witnesses in the network can be enforced to submit a task in every time slot $T$ in a round robin manner.

\subsection{Result Verification} \label{sec::res_verfication}
EntrapNet has two result verification processes. One occurs in constructing the fishing-task repository and the other in verifying the outsourced computation. 

In the phase of constructing the fishing-task repository, $N$ witnesses execute the same computing task and output results, vector $Y_i$ (the outcome from the $i$-th witness with $i=1,2,\cdots, N$). Considering the randomness inside the task script and hardware inaccuracy of the computing device, the results from multiple executions are defined to be aligned if any pair of the normalized results lie within a margin, $\Delta_{val}$, under the Euclidean norm, i.e.,
$$ \Big \Vert \frac{Y_i}{||Y_i||} - \frac{Y_j}{||Y_j||} \Big \Vert \leq \Delta_{val},$$
for $i\neq j$.
This metric can be easily generalized to weighted norms (e.g., infinity norm) or other notion of distance. If all results are proved to be aligned, the mean value $$\bar{Y}= \frac{1}{N}\sum_{i=1}^N Y_i,$$ along with the task script, the network verification proof and a user-specific key, will be used to generate a hash-based abstract and recorded on blockchain. This indicates that this fishing task is verified by the network and can be used later for entrapment. 

Another verification process corresponds to ensuring that the fishing-task results $Y_f$ from the provider is correct. This occurs when a fishing task is complete by a provider and its results are fed back to the officer.  The officer then checks the following:
\begin{equation}\label{equ:res_verification}
\frac{||Y_f - \bar{Y} ||}{||\bar{Y}||} \leq \Delta_{ver}.
\end{equation}
If the above inequality is violated, the officer will open a case to claim the misconduct of the provider. Note that in applications the margins $\Delta_{val}$ and $\Delta_{ver}$ need to be carefully tuned determined in prior by the network. In addition, the aforementioned verification can be performed on a dedicated part of the result vector rather than the whole vector $Y_i$ when $Y_i$ has a large dimension. 

Note that how tasks (such as machine learning tasks) can be handled by this system as the output may be non-deterministic and/or discrete (e.g., a classification task) will be a topic for the future research. 

\subsection{Judgement}

If a computing outcome is identified to be faulty, the officer will initiate an appeal by submitting necessary information to the verification smart contract. The information, as an input to the smart contract, includes the provider's results $Y_f$ and the four fishing-task information defined in Section \ref{sec::fishing_task}. That is, 
\begin{enumerate}
    \item task script;
    \item computing results;
    \item network verification proof;
    \item a user-specific key (e.g., a private key or a field like Bitcoin nonce).
\end{enumerate}
Then the judges, a subset or all of the nodes in the network, will run this verification smart  contract and hence rule  on  whether  the  officer's appeal is valid. The verification smart contract will execute the followings in order.
\begin{enumerate}
    \item \textit{Fishing-task verification}: Using the fishing-task information submitted by the officer, the contract will first confirm if this fishing task is a network-verified task. In particular, the contract will generate the abstract from the fishing-task information and compare it with the one stored on the blockchain which was submitted by the officer in the phase of building his fishing-task repository (Section \ref{sec::fishing_task});
    \item \textit{Result verification}: The contract will verify the results $Y_f$ according to  (\ref{equ:res_verification}); 
    \item \textit{Incentive allocation}: If the provider's result $Y_f$ is proved to be incorrect, the contract will reward  the officer from the incentive  pool,  and  have the provider's deposit forfeited and contributed to the incentive pool.
\end{enumerate}

\subsection{Incentive Protocol}\label{incentive protocol}
In this section, we discuss the EntrapNet incentive protocol which could effectively discourage the provider misconduct.  The proposed incentive protocol \textcolor[rgb]{0,0,0}{is controlled by smart contract on blockchain,  and} consists of three financial components, which are presented in order as follows. Due to anonymity of all participants, we need to consider Sybil attacks in our incentive protocol, in which the attacker subverts the network by creating a large number of pseudonymous identities. In our paper,  we assume that an attacker may create as many identities as he likes. The EntrapNet resists such attacks via its incentive protocol.

\subsubsection{Cost of generating fishing tasks}
When outsourcing a computing task, a user must fairly compensate the provider for computing such a task. This compensation should be at least higher than the cost of the provider for doing this computation job. As described in Section \ref{sec::fishing_task}, to generate a fishing task, a task is always assigned to a set of reputable and trustable users (a.k.a. witness) rather than, like a regular task, assigned to only one provider in the network. Since there exist only a few witnesses in the network, if the compensation for computing a fishing task is low, a Sybil attack by consistently submitting a large number of fishing tasks in a period from a number of anonymous participants will easily fully occupy all witnesses computing resources, leading to the situation in which the witnesses have no capacity for executing its normal jobs, e.g., verifying transactions on blockchain as used in many directed acyclic graph (DAG)-based blockchain protocols \cite{Byteball},\cite{Canonchain}. As a result, EntrapNet imposes high cost on generating a fishing task. That is, the user needs to pay much more for having a fishing task that requires verification from the network than having a regular task. Note, however, that if the reward for reporting a malicious provider is much higher than the cost of preparing a fishing task, the willingness of generating fishing tasks from users should remain. 

\subsubsection{Deposit of Officer and Provider}
EntrapNet requires deposit from providers to discourage their misconduct, and from officers to thwart Sybil attacks. Specifically,  if a provider misconduct is detected by the officer, the deposit $D$ from the provider, as penalty, will be forfeited and contributed to the EntrapNet's incentive pool, which is owned and maintained by the EntrapNet. The payout as a reward to the officer is taken from the incentive pool. To balance the incentive pool and guarantee the EntrapNet's effectiveness on security,  an adaptive mechanism on deposit $D$ is imperative. At this point, the design could be quite different case by case.  Nevertheless, we will provide a basic guidance in the next section. 

On the other hand, in order to avoid Sybil attack, a small amount of deposit $\tilde{D}$ from officer who reports a misconduct is required. This small deposit, however, must be enough to pay for the cost of judges (e.g., in Ethereum, at least the gas cost for running the verification smart contract) to perform the result verification. We assume $\tilde{D} << D$.

\subsubsection{Reward of reporting misconduct} \label{sec:: reward_misc}
First of all, the reward $R$ to the officer for successfully reporting a misconduct needs to be derived from the target security level of the network, \textcolor[rgb]{0,0,0}{which can be measured by the expected probability for a provider being assigned with a fishing task $p$. Since the expected probability $p$ will change over time, the reward $R$ is also dynamic and will be recalculated periodically in the system.}
Then, the reward $R_t$ at time $t$ is determined by 
\begin{equation}\label{equ::rho}
R_t = f(p_t),
\end{equation}
\textcolor[rgb]{0,0,0}{where $p_t$ is the expected probability $p$ at time t, and $f$ is a continuous, invertible and non-decreasing function mapping the security level to a reward. Over the long run, the function $f$ can be learned/approximated from the data set $(R_t,p_t)$, where the expected probability $p_t$ can be calculated by using the arrival rate of fishing tasks and regular tasks in the network (details are given in Section \ref{sec::opt}), and the data set can be obtained by adjusting reward $R$ in the EntrapNet and observing the real-time throughput of tasks.}

Next, we discuss the \textcolor[rgb]{0,0,0}{non-negative} property of the incentive pool, a fact that is necessary to guarantee the long-term economic operation of EntrapNet. Assume that in time slot $t$ the expected number of true misconduct reports is $n_t$. Then the averaged deposit forfeited and contributed to the incentive pool is $L_t = n_t D_t$ where $D_t$ represents the required deposit in time slot $t$, while the averaged reward taken from the pool to the officers is $G_t = n_tR_t$. EntrapNet balances the incentive pool by maintaining the following conservation law,
$$\lim_{T\rightarrow \infty} \left( \sum_{t=1}^{T}L_t - \sum_{t=1}^TG_t \right) > 0.$$
Note that this equality does not imply \textcolor[rgb]{0,0,0}{$D_t > R_t$} for $\forall t$. Given a target security \textcolor[rgb]{0,0,0}{$p_t$ }(thus $R_t$ in (\ref{equ::rho})) and $n_t$ reported misconducts, this equality provides a basic guidance on dynamically setting the deposit $D_t$.
%Following the guidance, one example of setting deposit can be the following: for a target security level $\rho_T$, we set reward $R_T = f(\rho_t)$ and then we count $n_T$ explicitly in period $T$. Then we set $D_{T+1}$ such that the above equality holds. 

\section{System Optimization} \label{sec::opt}
In the preceding section, we have introduced the architecture and key components of EntrapNet. Although there exist several hyper-parameters to tune in applications, the most important one is the rate of submitting fishing tasks for a given arrival rate of the regular tasks. The higher the rate is, the more securely the network performs. However, higher rate implies that the network-wise overhead of computing the fishing tasks can be overwhelming as computing fishing tasks is completely waste of network resources. Therefore, the tradeoff between network security and efficiency needs to be carefully studied. In this section, we optimize this tradeoff by providing solutions to the following problem: \\

\textit{For a given metric on network security and efficiency, what is the optimal rate of submitting fishing tasks corresponding to the arrival rate of regular tasks in the network?}\\

%In essence, we aim to find the optimal security level (defined in Section \ref{sec:: reward_misc}) to maximize the utility/performance of the network. 
%Specifically, if the rate of sending fishing tasks is high, the network is more secure but the network-wise overhead of computing the fishing tasks can be overwhelming. In contract, if the rate is low, the network is vulnerable. I
In what follows, we first propose a mathematical framework to formulate this problem, and then provide analytical solutions to the optimization problem. We present all proofs in Appendix.

\subsection{Problem formulation}
Let us first introduce some assumptions and notations as follows. \textcolor[rgb]{0,0,0}{Here we define $X$ and $Y$ be the respective number of fishing tasks and regular tasks in the scheduling queue in one time interval.}
\begin{assumption}\label{poisson}
The number of both fishing tasks and regular tasks submitted to the network within a fixed time interval follows Poisson distribution $Pois(x;\lambda_x)$ and $Pois(y;\lambda_y)$, respectively. 
\end{assumption}
This assumption follows the fact that the sum of many small \textit{independent} arrival processes tends to be close to Poisson even if the small processes are not, where in our settings the action of submitting a (regular or fishing) task from each user can be assumed to be independent. 

\begin{assumption}
When a provider is assigned with a task, it executes the task for sure.
\end{assumption}
As EntrapNet ensures that a provider cannot distinguish tasks (a fishing task or not),  this assumption indicates that a malicious provider will always execute a fishing task, if assigned, and thus be certainly detected by the result verification process (in Section \ref{sec::res_verfication}).

 Under the above two assumptions and following the EntrapNet's uniform randomly task scheduler (in Section \ref{sec::task_schedular}), the expected probability for a provider being assigned with a fishing task is given by\footnote{\textcolor[rgb]{0,0,0}{This ratio is inspired by the fact that if the processes of $X$ and $Y$ follow Poisson distribution, i.e., $X \sim Pois(x, \lambda_x)$ and $Y \sim Pois(y,\lambda_y)$, then $X+Y \sim Pois(x+y, \lambda_x+\lambda_y)$.}}
\begin{equation}\label{equ::p}
p_{\lambda_y}(\lambda_x) = \mathbb{E}\left[\frac{X}{X+Y+1}\right].
\end{equation}
Recall that in Assumption \ref{ass:onetask} EntrapNet always imposes one task in every scheduling interval. Therefore, the total number of tasks in one interval is $X+Y+1$. Otherwise the denominator could be ill-defined if $X=Y=0$. Furthermore, one can see that $p_{\lambda_y}(\lambda_x)$ implies the security level of the network. However, although both random variables $x$ and $y$ follow Poisson process, a closed-form characterization on $p_{\lambda_y}(\lambda_x)$ is unknown, a fact that becomes the utmost difficulty in solving the proposed optimization (to be seen later). 
In addition, the fishing task rate $\lambda_x$, which can be directly used to measure the network overhead, should be tuned within an appropriate regime.  Without loss of much generality, we assume this regime to be a convex set $\mathcal{C}$. 

Next, similar to the well-studied networking science, a utility function that measures the network security and efficiency is considered in our framework. This utility is a function of the expected probability $p_{\lambda_y}(\lambda_x)$ (corresponding to the network security) and the fishing task rate $\lambda_x$ (corresponding to the network efficiency). Putting the above together, the optimization of finding an optimal fishing task rate $\lambda_x$ is formulated as 

\begin{equation}
{\bf OP}: \quad \mu = \max_{\lambda_x\in \mathcal{C}} \quad U\big (p_{\lambda_y}(\lambda_x), \lambda_x \big),
\end{equation}
where $U(a,b)$ is assumed to be an $L$-Lipschitz continuous utility function that is concave in $(a, b)$. The $L$-Lipschitz property implies 
\begin{equation} \label{eqa:L_lipschitz}
|U\big (a,b \big) - U\big (a',b'  \big)| \leq L \cdot ||[a,b]-[a',b']||,
\end{equation}
where $||\cdot||$ is the Euclidean norm. In general, solving {\bf OP} is cumbersome as it is nontrivial to explicitly characterize $p_{\lambda_y}(\lambda_x)$. In addition, even if $p_{\lambda_y}(\lambda_x)$ can be approximated by a closed form function in $\lambda_x$, we wish such a function to be concave in $\lambda_x$ as well. Otherwise, the objective function in {\bf OP} might be neither convex nor concave, resulting in an non-convex optimization whose global optimal solution is unknown in general. 

\subsection{Asymptotic Optimal Solutions to  {\bf OP}}
Since the optimization {\bf OP} cannot be easily characterized and solved by standard optimization tools, our idea is to construct a new \textit{convex} optimization which asymptotically converges to OP. Then we characterize the gap between the two optimizations, which is proved to be useful in engineering (when dealing with non-asymptotic scenarios). 

Firstly, we introduce the following optimization {\bf OP1} which has an explicit characterization on the first argument of the utility function. 

\begin{equation}
\begin{split}
{\bf OP1 }: \quad \mu_1 =  &\max_{\lambda_x\in \mathcal{C}} \quad U\big (\bar{p}_{\lambda_y}(\lambda_x), \lambda_x \big)\\
s.t. \quad &\bar{p}_{\lambda_y}(\lambda_x) = 1- \frac{1}{1+\alpha(\lambda_x, \lambda_y)},\\
\end{split}
\label{op1}
\end{equation}
where $\alpha(\lambda_x, \lambda_y) = \frac{\lambda_x}{\lambda_y}(1-e^{-\lambda_y})$.
Notice that $\bar{p}_{\lambda_y}(\lambda_x)$ is concave in $\lambda_x$. Then the following lemma holds. 
\begin{lemma}\label{lem::concavity}
Consider a convex set $\mathcal{C}$ and a concave utility function $U(a,b)$ in $(a,b)$ in {\bf OP1}. If $U(a,b)$ is nondecreasing in $a$ for $\forall b$, the optimization {\bf OP1} is concave in $\lambda_x$.
\end{lemma}

This Lemma implies that in general it is straightforward to solve ${\bf OP1}$ due to its concavity. However,  how to find a (near)-optimal solution to the original problem ${\bf OP }$ from the solution to {\bf OP1}? The following lemma shows that the well characterized $\bar{p}_{\lambda_y}(\lambda_x)$ is in fact an upper bound on $p_{\lambda_y}(\lambda_x)$ in ${\bf OP}$, and converges to $p_{\lambda_y}(\lambda_x)$ as $\lambda_y$ increases. Based on this result, we further prove that ${\bf OP1}$ provides an asymptotic optimal solution to the original optimization ${\bf OP}$.

%\begin{definition}
%For given $\lambda_y$ and $\epsilon>0$, {\bf OP} is said to be solved with $\epsilon$ accuracy if there exists $\lambda_x$ such that $|\mu-\mu_1|<\epsilon$.
%\end{definition}

\begin{lemma}  \label{lem::bounds}
Let $X$ and $Y$ be real random variables following Possion distribution $Pois(x; \lambda_x)$ and $Pois(y; \lambda_y)$, respectively. Then the following inequalities hold:
$$lb_{\lambda_y}(\lambda_x)\leq \mathbb{E}\left[\frac{X}{X+Y+1}\right]  \leq ub_{\lambda_y}(\lambda_x).$$
The upper and lower bounds are respectively given by 
\begin{equation}
ub_{\lambda_y}(\lambda_x) = 1- \frac{1}{1+\alpha(\lambda_x, \lambda_y)},
\end{equation}
and 
\begin{equation}
%lb_{\lambda_y}(\lambda_x) = \alpha(\lambda_x, \lambda_y)\big[ 1 - (1+\lambda_x) \sum_{j=1}^\infty \frac{(j-1)!}{\lambda_y^j}\big],
lb_{\lambda_y}(\lambda_x) = \alpha(\lambda_x, \lambda_y) -\frac{\lambda_x (1+\lambda_x)}{\lambda_y} \sum_{j=1}^\infty \frac{(j-1)!}{\lambda_y^j},
\end{equation}
where $\alpha(\lambda_x, \lambda_y) = \frac{\lambda_x}{\lambda_y}(1-e^{-\lambda_y})$.
\end{lemma}

 \begin{figure*}
\begin{center}
\includegraphics[scale=0.6]{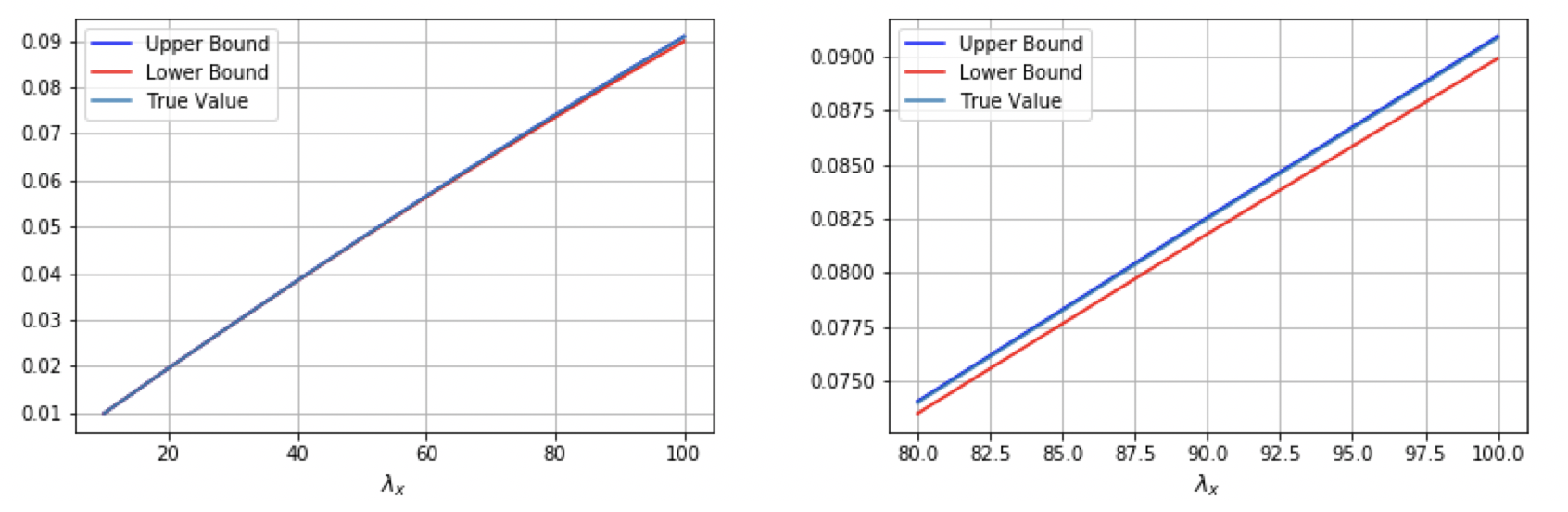}
\caption{Bound evaluation: $\lambda_y = 1000$. Left: upper bound, lower bound and true value; Right: zoom-in on the regime of interest (i.e., $\lambda_x\in [80,100]$).}
\label{fig:gap01}
\end{center}
\end{figure*}

\begin{figure}
\begin{center}
\includegraphics[scale=0.6]{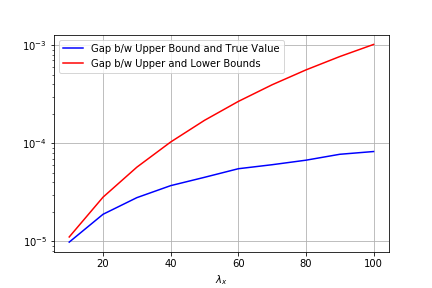}
\caption{Gap evaluation: : $\lambda_y = 1000$. }
\label{fig:gap02}
\end{center}
\end{figure}

\begin{remark}
For a fixed $\lambda_y$, this lemma explicitly characterizes a pair of upper and lower bounds on $p_{\lambda_y}(\lambda_x)$ in {\bf OP}.  As shown in Fig. \ref{fig:gap01} and Fig. \ref{fig:gap02}, these bounds are nearly tight. As it is shown in the zoomed figure (Fig. \ref{fig:gap01}, right) and Fig. \ref{fig:gap02}, the gap between the upper bound and the true value falls in the regime less than $10^{-4}$ (almost tight !). This observation motivates us to  use upper bound to characterize $p_{\lambda_y}(\lambda_x)$ in {\bf OP}. The other motivation comes from the concavity of this upper bound. 
\end{remark}

\begin{remark} \label{remark::asym}
Assume a fixed $\lambda_x^{max}>0$, for any $\lambda_x \in (0,\lambda_x^{max}]$, one can easily check the following facts:
$$\lim_{\lambda_y \rightarrow \infty} lb_{\lambda_y}(\lambda_x) = 0, $$
and 
$$\lim_{\lambda_y \rightarrow \infty} ub_{\lambda_y}(\lambda_x) = 0.$$
\end{remark}

Next, for a given $\lambda_y$, we define the \textit{maximal gap} as
$$\rho_{\lambda_y} = \max_{\lambda_x \in (0,\lambda_x^{max}]}  \quad ub_{\lambda_y}(\lambda_x)  - lb_{\lambda_y}(\lambda_x). $$ 
Note that, since $ub_{\lambda_y}(\lambda_x)$ and $lb_{\lambda_y}(\lambda_x)$ are well characterized,  the value $\rho_{\lambda_y}$ can be easily obtained by numerically searching $\lambda_x \in (0,\lambda_x^{max}]$.

Now, we present the main result as follows.
\begin{theorem} \label{thm::main}
Let $X$ and $Y$ be real random variables following Possion distribution $Pois(x; \lambda_x)$ and $Pois(y; \lambda_y)$, respectively. Denote $\mu$ and $\mu_1$ as the optimal objective value of the optimizations ${\bf OP}$ and ${\bf OP1}$, respectively. $L$ is the Lipschitz coefficient of the utility function $U(\cdot, \cdot)$ in ${\bf OP}$ and ${\bf OP1}$. Denote $\lambda_x^{max}$ as the maximal submission rate of the fishing tasks in the network. Then for a fixed $\lambda_y$ the following inequality holds, 
$$|\mu-\mu_1| \leq L\cdot \rho_{\lambda_y},$$
where  
$\lim_{\lambda_y \rightarrow \infty} \rho_{\lambda_y}= 0.$
\end{theorem}

This theorem estimates the accuracy of using {\bf OP1} to approximate the original optimization {\bf OP}.  That is, by solving the convex optimization {\bf OP1}, one obtains a solution $\lambda_x$ which leads to an objective function value lying within a margin $L\cdot \rho_{\lambda_y}$ to {\bf OP}. Since this margin is shrinking to zero as the arrival rate of regular tasks $\lambda_y $ increases, the solution is asymptotically optimal. 

\subsection{Example} 
In this section, we present a concrete example of optimizing the proposed EntrapNet  \textcolor[rgb]{0,0,0}{based on the Ethereum Blockchain. That is, the EntrapNet is assumed to be built on Ethereum and each part of EntrapNet (in Section \ref{sec::arch}) is realized by the Ethereum smart contract. For now, Ethereum, a decentralized open source blockchain featuring smart contract functionality is the second largest cryptocurrency platform by market capitalization. Next, we set a few parameters for our example. Let $p_{\lambda_x}^{min}$ be the minimal expected probability for a provider being assigned with a fishing task, $\lambda_x^{max}$ be the maximal rate of submitting fishing tasks in the network, $D$ be the deposit required from a provider for computing a task, and $R$ be the dynamic reward to an officer for successfully detecting a provider misconduct. Considering the logarithm cost criteria, which is frequently used in utility optimizations, the objective function of interest is proposed as follows:}

$$U\big (p_{\lambda_y}(\lambda_x), \lambda_x \big) = \log(1-\frac{p_{\lambda_x}^{min}}{p_{\lambda_y}(\lambda_x)}) - \big( -c_1\cdot  \log(1-\frac{\lambda_x}{\lambda_x^{max}})\big)\\$$
$$+c_2\cdot\log(1-\frac{R(\lambda_x)}{D}),$$
\textcolor[rgb]{0,0,0}{where the first term $\log(1-\frac{p_{\lambda_x}^{min}}{p_{\lambda_y}(\lambda_x)})$ measures the network security level. Since $p_{\lambda_y}(\lambda_x)$ is an increasing function of variable $\lambda_x$, it is easy to see that the first term is an increasing function of $\lambda_x$. That is, a higher fishing task arrival rate $\lambda_x$ which indicates a higher security level results in a larger value of the first term; the second term $-c_1\cdot \log(1-\frac{\lambda_x}{\lambda_x^{max}})$ is assumed to be the network-wise cost of sending fishing tasks, which is an increasing function of $\lambda_x$. That is, a smaller $\lambda_x$ indicates a lower cost of the network; the third term $c_2\cdot\log(1-\frac{R}{D})$ guarantees the profitability of EntrapNet, which ensures the long-term operation of the protocol, and is an decrasing function of $\lambda_x$. Parameter $c_1$ and $c_2$ are used to balance the three terms. In addition, the dynamic reward $R(\lambda_x)$ is a function of the expected probability $p_{\lambda_y}(\lambda_x)$ (see Section \ref{sec:: reward_misc}). In this example this reward function that fits the experimental data is modeled as 
\begin{equation}\label{reward}
R(\lambda_x) = \frac{105}{1+e^{-(100p_{\lambda_y}(\lambda_x)-5)}},
\end{equation}
Putting the above together, the optimization problem {\bf OP} is given by}
\begin{equation}\label{utility}
\begin{split}
{\bf OP }: \quad \mu =  &\max_{\lambda_x} \quad U\big (p_{\lambda_y}(\lambda_x), \lambda_x \big)\\
s.t. \quad& R(\lambda_x) = \frac{105}{1+e^{-(100p_{\lambda_y}(\lambda_x)-5)}},\\
 & \lambda_x \in (0,\lambda_x^{max}].\\
\end{split}
\end{equation}

\textcolor[rgb]{0,0,0}{
Then the corresponding {\bf OP 1} in (\ref{op1}) is given by }
\begin{equation}\label{utility}
\begin{split}
{\bf OP 1 }: \quad \mu_1 =  &\max_{\lambda_x} \quad U\big (\bar{p}_{\lambda_y}(\lambda_x), \lambda_x \big)\\
s.t. \quad& R(\lambda_x) = \frac{105}{1+e^{-(100\bar{p}_{\lambda_y}(\lambda_x)-5)}},\\
 & \lambda_x \in (0,\lambda_x^{max}].\\
\end{split}
\end{equation}

\textcolor[rgb]{0,0,0}{
In this example, EntrapNet employs a deposit $D = 100$ (e.g., in US dollars), and the balancing parameter $c_1 = 1, c_2=0.1$. We choose a small $c_2$ because  the first two terms (representing the tradeoff between the security and the efficiency of EntrapNet) in the objective function are more relevant.  In addition, since the EntrapNet is assumed to be built on Ethereum, we should consider Ethereum's throughput limitation when choosing the maximum task rate. The maximum throughput of Ethereum in practice is nearly 20 transactions per second\cite{Eth throughput}, which means 1200 transactions per slot ($1$ slot = $1$ minute). Considering the basic efficiency of the network, we set the maximal rate of submitting fishing transactions in this example to be $\lambda_x^{max} = 120$ transactions per slot (about 10$\%$ overhead). Also, we set the arrival rate of regular transactions to be $\lambda_y = 1000$  transactions per slot. Therefore, the maximal throughput is $1120$ transactions per slot that satisfies the throughput limitation of Ethereum blockchain (i.e., 1200 transactions per slot). Next, we assume a minimum security level $p_{\lambda_x}^{min}= 0.01$. According to this setting, one can derive an $L$-Lipschitz coefficient $L = 1$ from the basis of the logarithm function property. Figure 5 shows the proposed utility function which is strictly convex w.r.t. the fishing-task rate $\lambda_x$.
Using the standard CVX optimization toolbox (a package for specifying and solving convex
programs)\cite{cvx_tool01,cvx_tool02}, the optimal solution to ${\bf OP1}$ that corresponds to the above optimization ${\bf OP}$ is $\lambda_x = 33.4$. The achieved objective value $\mu_1 = -0.71$. The resulting margin characterized in Theorem \ref{thm::main} is $L\cdot p_{\lambda_y}(\lambda_x)= 0.000476$, corresponding to an accuracy of using ${\bf OP1}$ to approximate the original optimization ${\bf OP}$  at least better than $$1-\frac{L\cdot \rho_y}{\max\lbrace \mu_1+L\cdot \rho_y, \quad \mu_1-L\cdot \rho_y\rbrace} = 99.93\%.$$}
\begin{figure}
\begin{center}
\includegraphics[scale=0.55]{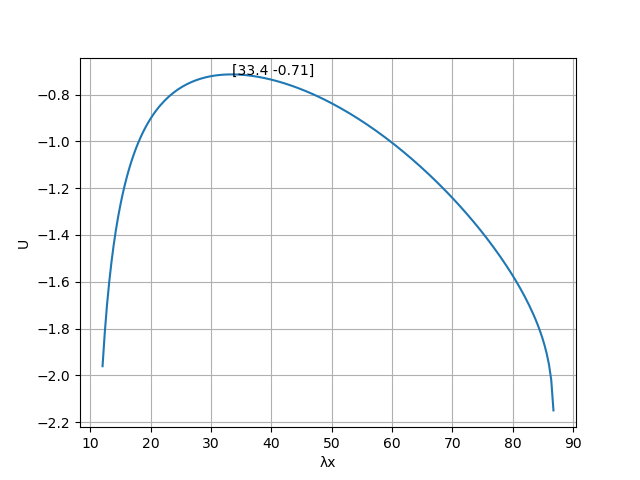}
\caption{Utility function with c1=1, c2=0.1, D=100. The optimal $\lambda_x = 33.4$, $\mu = - 0.71$. }
\label{fig:optimal U}
\end{center}
\end{figure}

\begin{figure}
\begin{center}
\includegraphics[scale=0.55]{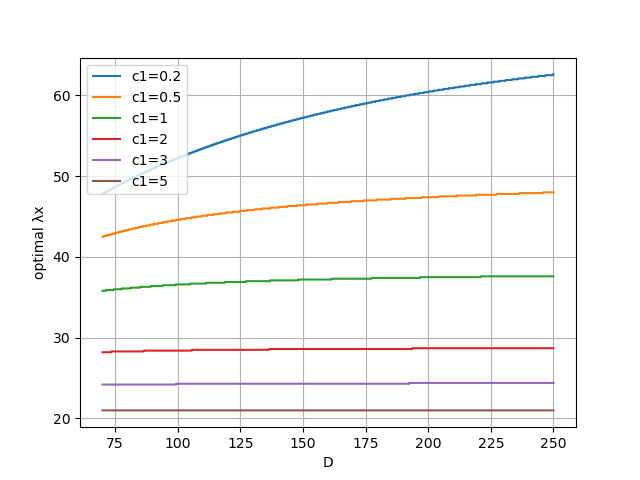}
\caption{Relationship between the deposit $D$ and the optimal $\lambda_x$ }
\label{fig:D}
\end{center}
\end{figure}

\textcolor[rgb]{0,0,0}{
In essence, the proposed utility model is balancing the security and the efficiency of the network.
According to the utility function (\ref{utility}), we see that different deposit value $D$ from providers results in different optimal fishing task rate $\lambda_x$. Figure \ref{fig:D} shows the relationship between the deposit $D$ and the optimal fishing task rate $\lambda_x$ with different balancing parameter $c_1$. Notice that The deposit $D$ can be also viewed as a measure of the network security. In particular, higher deposit $D$ means higher risk for cheating, implying a much more secure network. On the other hand, a larger $\lambda_x$ implies a lower efficient network when $\lambda_y$ is fixed. As shown in Figure  \ref{fig:D}, as the deposit $D$ increases, the optimal $\lambda_x$ increases very slowly. This implies that when we enhance the security level of the network, we do not lose the network efficiency quite much. That is, the utility function (\ref{utility}) in the this example is well designed.
}

\begin{figure}
\begin{center}
\includegraphics[scale=0.55]{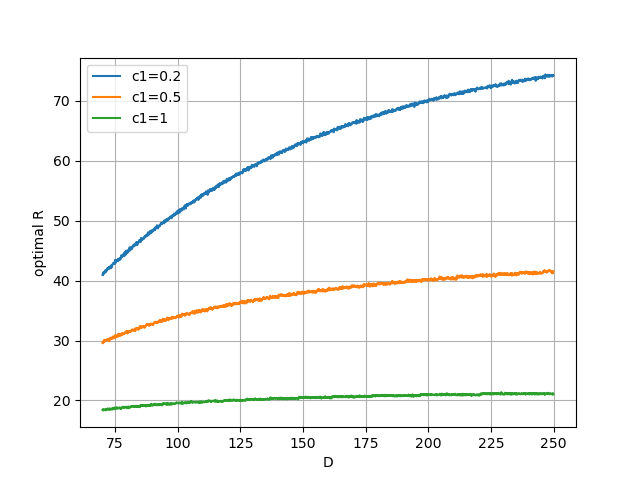}
\caption{relationship between deposit $D$ and optimal reward $R$ }
\label{fig:R}
\end{center}
\end{figure}
\textcolor[rgb]{0,0,0}{Next, we discuss how to set the the reward $R$ to navigate the network to achieve the optimal fishing-task rate $\lambda_x$. Figure \ref{fig:R} represents the relationship between the deposit $D$ and the optimal reward $R$ (the $R$ value in (\ref{reward}) that corresponds to the optimal fishing-task rate $\lambda_x$) on different balancing parameter $c_1$. In order to obtain the optimal $R$ value, the approach to evaluate $p_{\lambda_y}(\lambda_x)$ in (\ref{reward}) is described as follows. We first solve {\bf OP 1} to obtain the optimal $\lambda_x$ w.r.t. the different deposit $D$ and the balancing parameter $c1$ and $c2$. Then according to Assumption \ref{poisson}, it is straightforward to obtain the expected probability $p_{\lambda_y}(\lambda_x)$ via numerical simulations. With the results in Figure \ref{fig:R} in hand, one can easily look up the optimal setting of reward $R$ that leads to the optimal fishing-task rate $\lambda_x$.}

\section{Conclusion} 
In this paper, we have proposed a blockchain-based computing verification protocol, called EntrapNet, which borrows the idea from the practice of criminal entrapment. \textcolor[rgb]{0,0,0}{The EntrapNet can complete the verification on outsourced computational tasks with very low cost.}  We have further proposed a mathematical framework to analyze EntrapNet with the objective to optimize the tradeoff between the network \textit{security} and \textit{efficiency}. EntrapNet is agnostic to the underlying blockchain and can be directly applied to any distributed computing network that requires outsourced computation to trustless parties. \textcolor[rgb]{0,0,0}{In the end, we have presented a simulation example of EntrapNet that is built on the Ethereum blockchain, and provided insights in how to optimize the network. In the future, more sophisticated blockchain-based schemes can be designed based on EntrapNet. For example, how the current approach can be applied to tasks that really need to be split amongst several providers and how would security and efficiency be calculated in that case?}

\section{appendix}
\subsection{Proof of Lemma \ref{lem::concavity}}
Firstly, in the optimization {\bf OP1} it is straightforward to check that $1- \frac{1}{1+\alpha(\lambda_x, \lambda_y)}$ is concave in $\lambda_x$. Next, since the utility function $U(a,b)$ is jointly concave in $(a,b)$, the utility function $U(a,b)$ is concave in $a$ for $\forall b$. Then, if $U(a,b)$ is nondecreasing in $a$ for $\forall b$, the claim directly follows from the composition rule of convex (or concave) functions (\cite{cvx_book} (Section 3.2.4, (3.10)).

\subsection{Proof of Lemma \ref{lem::bounds}}
Due to the page limit, the proof is omitted. 

\subsection{Proof of Theorem \ref{thm::main}}

First of all, we prove the following inequality on functions $f_1, f_2: R^n \rightarrow R$ with the domain defined in a compact set.
\begin{equation}\label{equ::max_ineq}
|\max_x f_1(x) -  \max_x f_2(x) | \leq \max_x |f_1(x) -  f_2(x)|.
\end{equation}
Assume $\max_x f_1(x) \geq \max_x f_2(x)$, then 
\begin{equation}
\begin{split}
&|\max_x f_1(x) -  \max_x f_2(x) | \\
=& \max_x f_1(x) -  \max_x f_2(x) \\
= & \max_x [f_2(x) + f_1(x) - f_2(x)] -  \max_x f_2(x) \\
\leq & \max_x f_2(x) + \max_x [f_1(x) - f_2(x)] -  \max_x f_2(x) \\
= & \max_x [f_1(x) - f_2(x)] \\
\leq & \max_x |f_1(x) - f_2(x)|. \\
\end{split}
\end{equation}
Following the same approach, it is straightforward to check that the above proof holds if assuming $\max_x f_2(x) \geq \max_x f_1(x)$. Thus, the inequality (\ref{equ::max_ineq}) is true. Now, we are ready to prove the theorem. Denote 
the objective value $U\big (\bar{p}_{\lambda_y}(\lambda_x), \lambda_x \big)$ as $U\big (ub_{\lambda_y}(\lambda_x), \lambda_x \big)$ with $p_{\lambda_y}(\lambda_x) = ub_{\lambda_y}(\lambda_x)$. Then,

\begin{equation} \label{equ::thm_proof_01}
\begin{split}
&|\mu - \mu_1|\\
=&\big| \max_{\lambda_x\in \mathcal{C}} \quad U\big (p_{\lambda_y}(\lambda_x), \lambda_x \big) - \max_{\lambda_x\in \mathcal{C}} \quad U\big (ub_{\lambda_y}(\lambda_x), \lambda_x \big) \big |\\
\leq & \max_{\lambda_x\in \mathcal{C}} \big| U\big (p_{\lambda_y}(\lambda_x), \lambda_x \big) -  U\big (ub_{\lambda_y}(\lambda_x), \lambda_x \big) \big |.\\
\end{split}
\end{equation}
Let $\lambda_x^*$ be the solution to the above optimization (last line in (\ref{equ::thm_proof_01})), then
\begin{equation} 
\begin{split}
 & \max_{\lambda_x\in \mathcal{C}} \big| U\big (p_{\lambda_y}(\lambda_x), \lambda_x \big) -  U\big (ub_{\lambda_y}(\lambda_x), \lambda_x \big) \big |\\
 = & \big| U\big (p_{\lambda_y}(\lambda_x^*), \lambda_x^* \big) -  U\big (ub_{\lambda_y}(\lambda_x^*), \lambda_x^* \big) \big |\\
\leq & \max_{\lambda_x\in \mathcal{C}} \big | U\big(p_{\lambda_y}(\lambda_x), \lambda_x^* \big) -  U\big (ub_{\lambda_y}(\lambda_x), \lambda_x^* \big) \big |\\
\overset{(a)}{\leq}  & \max_{\lambda_x\in \mathcal{C}}L \Big |\Big | [p_{\lambda_y}(\lambda_x),  \lambda_x^*] - [ub_{\lambda_y}(\lambda_x),  \lambda_x^*]  \Big |\Big | \\
=  & \max_{\lambda_x\in \mathcal{C}}L  ( |p_{\lambda_y}(\lambda_x)-ub_{\lambda_y}(\lambda_x)|^2 - |\lambda_x^*-  \lambda_x^*|^2 )^{1/2}\\
= & \max_{\lambda_x\in \mathcal{C}} L | p_{\lambda_y}(\lambda_x)  - ub_{\lambda_y}(\lambda_x) | \\
\leq & L \max_{\lambda_x\in \mathcal{C}} | lb_{\lambda_y}(\lambda_x)  - ub_{\lambda_y}(\lambda_x) | \\
= & L \rho_{\lambda_y} .
\end{split}
\end{equation}
where step (a) follows from the $L$-Lipschitz property of the utility function.
Next, we show that $\lim_{\lambda_y \rightarrow \infty} \rho_{\lambda_y}= 0$. Specifically,
\begin{equation} 
\begin{split}
\lim_{\lambda_y \rightarrow \infty} \rho_{\lambda_y}&= \lim_{\lambda_y\rightarrow \infty} \max_{\lambda_x\in \mathcal{C}} | lb_{\lambda_y}(\lambda_x)  - ub_{\lambda_y}(\lambda_x) |\\
&= \lim_{\lambda_y\rightarrow \infty} \max_{\lambda_x\in \mathcal{C}} \bigg( ub_{\lambda_y}(\lambda_x)  - lb_{\lambda_y}(\lambda_x)\bigg) \\
&\leq \lim_{\lambda_y\rightarrow \infty} \max_{\lambda_x\in \mathcal{C}} ub_{\lambda_y}(\lambda_x) \\
&\leq \lim_{\lambda_y\rightarrow \infty} ub_{\lambda_y}(\lambda_x^{max}) \\
&=0\\
\end{split}
\end{equation}
where the last step follows from Remark \ref{remark::asym}. The proof is complete.


\begin{thebibliography}{00}
\bibitem{canonchain} Canonchain project whitepaper, http://www.canonchain.com/.
\bibitem{golem} Golem project whitepaper, https://golem.network/home/.
\bibitem{sonm} Sonm project whitepaper, https://sonm.com/.
\bibitem{zcash} Zcash project whitepaper. http://z.cash/. 
\bibitem{multiServer01} R. Canetti, B. Riva. and G. Rothblum. ``Practical delegation of computation using multiple servers''. In \textit{Proceedings of the 18th ACM conference on computer and communications security}, 2011, pp. 445-454.

\bibitem{multiServer02} M. Castro, and B. Liskov. ``Practical Byzantine fault tolerance and proactive recovery''. \textit{ACM Trans. on Comp. Sys}, vol. 20, no. 4, 398-461, 2002.

\bibitem{multiServer03} D. Malkhi and M. Reiter, ``Byzantine quorum systems'', \textit{Distributed Computing} vol. 11, no. 4, pp. 203-213, 1998.

\bibitem{trustHardware} A.R.Sadeghi, T. Schneider and M. Winandy, ``Token-based cloud computing: Secure outsourcing of data and arbitrary computations with lower latency''. In \textit{Proceedings of TRUST}, 2010.

\bibitem{Bootstrapping} B. Parno, J.M. McCune, and A. Perrig,``Bootstrapping Trust in Modern Computers''. Springer Science \& Business Media, 2011. 

\bibitem{Pinocchio} B. Parno and C. Gentry, ``Pinocchio: Nearly practical verifiable computation'', \textit{IEEE Symposium on Security and Privacy}, Oakland, 2013, pp. 238-252.

\bibitem{TinyRAM} E. Ben-Sasson, A. Chiesa, D. Genkin, E. Tromer, and M. Virza, ``SNARKs for C: verifying program executions succinctly and in zero knowledge''. In \textit{Advances in Cryptology-CRYPTO}, pp. 90-108, 2013.

\bibitem{Pantry} B. Braun, A. J. Feldman, Z. Ren, S. Setty, A. J. Blumberg, and M. Walfish. ``Verifying computations with state''. In \textit{Proceedings of SOSP}, Nov. 2013.

\bibitem{Buffet} R. S. Wahby, S. Setty, Z. Ren, A. J. Blumberg and M. Walfish. ``Efficient RAM and control flow in verifiable outsourced computation.'' In \textit{Proceedings of NDSS}, Feb. 2015.

\bibitem{survey_walfish} M. Walfish and A. J. Blumberg, ``Verifying computations without reexecuting them.'' \textit{Communications of the ACM} 58.2, pp. 74-84, 2015.

\bibitem{wiki::entrap} Wikipedia, ``https://en.wikipedia.org/wiki/Entrapment''.
\bibitem{TrueBit} J. Teutsch and C. Reitwiessner, ``A scalable verification solution for blockchains'', \textit{online}, Mar 2017.
\bibitem{biometric} Blanton, Marina, Yihua Zhang, and Keith B. Frikken. "Secure and verifiable outsourcing of large-scale biometric computations." ACM Transactions on Information and System Security (TISSEC) 16.3 (2013): 1-33.
\bibitem{Byteball} A. Churyumov, ``Byteball: A Decentralized System for Storage and Transfer of Value'' https://byteball. org/Byteball.pdf, 2016
\bibitem{Canonchain} L. Zhang, ``Consensus and Security in Canonchain'', \textit{online}, 2019.
\bibitem{JAAS} C. D. Coath, R. C. J. Steele and W. Fred Lunnon, ``Statistical bias in isotope ratios'', \textit{Journal of Analytical Atomic Spectrometry}, 28 (1), pp. 52-58, 2013.

\bibitem{PoS} I. Bentov, A. Gabizon, and A. Mizrahi. "Cryptocurrencies without proof of work." In International Conference on Financial Cryptography and Data Security, pp. 142-157. Springer, Berlin, Heidelberg, 2016.
\bibitem{Eth throughput}L. M. Bach, B. Mihaljevic and M. Zagar, "Comparative analysis of blockchain consensus algorithms", 41st International Convention on Information and Communication Technology, Electronics and Microelectronics (MIPRO), 2018.
\bibitem{cvx_tool01} M. Grant and S. Boyd, ``Cvx: Matlab software for disciplined convex programming'', version 2.0 beta,” 2013.
\bibitem{cvx_tool02}M. Grant and S. Boyd, ``Graph implementations for nonsmooth convex programs, recent
advances in learning and control,'' Lecture Notes in Control and Information Sciences, Springer, pp. 95-110, 2008.
\bibitem{cvx_book} S. Boyd and L. Vandenberghe, ``Convex Optimization'', Cambridge, U.K.: Cambridge Univ. Press, 2004.
\end{thebibliography}
\end{document}